\documentclass[1p]{elsarticle}

\usepackage{lineno}
\usepackage{graphicx}
\usepackage{algorithm}
\usepackage{algorithmic}
\usepackage{color}
\usepackage{amsmath}
\usepackage{amssymb}
\usepackage{multirow}

\modulolinenumbers[10]

\begin{document}

\begin{frontmatter}

\title{Dismantle a network efficiently during the entire process by a compound algorithm}

\author[cauc]{Shao-Meng Qin} \ead{qsminside@163.com}
\address[cauc]{College of Science, Civil Aviation University of China, Tianjin 300300, China}

\begin{abstract}
The dismantling network problem only asks the minimal vertex set of a graph after removing which the remaining graph will break into connected components of sub-extensive size, but we should also consider the efficiency of intermediate states during the entire dismantling process, which is measured by the general performance $R$ in this paper.
In order to improve the general performance of the belief-propagation decimation (BPD) algorithm, we implement a compound algorithm (CA) mixing the BPD algorithm and the node explosive percolation (NEP) algorithm.
In this CA, the NEP algorithm will rearrange and optimize the head part of a dismantling sequence of vertices which is given by the BPD algorithm firstly.
Two ancestor algorithms are connected at the joint point where the general performance can be optimized.
The numerical computations on Erd\"os-R\'enyi graphs, random-regular graphs, and scale-free graphs with various average degree prove the CA inherits the advantage of its two ancestors:
It can dismantle a graph to small pieces as quickly as the BPD algorithm, and it is with the efficiency of the NEP algorithm during the entire dismantling process.
By studying how the joint point of two ancestor algorithms affects the value of $R$, we find that a wise joint point is where the BPD algorithm breaks the original graph to subgraphs with the size just smaller than the $1\%$ of the original one.
We refer the CA with this settled joint point as the fast CA.
The fast CA overcomes a drawback in the original CA that the original CA is in the higher complexity class than the NEP algorithm.
The numerical computations on random graph ensembles with the size from $2^{10}$ to $2^{19}$ reveal that the fast CA is in the same complexity class with the BPD algorithm.
The computation on some real-world instances also exhibits that using the fast CA to optimize the intermediate process of a dismantling algorithm is an effective approach indeed.
\end{abstract}

\begin{keyword}
dismantling network, feedback vertex set, node explosive percolation
\MSC[2010] 05C38 \sep 05C40 \sep  05C80 \sep 05C85 \sep 68R10 \sep 90B15
%05C38 Paths and cycles
%05C40 Connectivity
%05C80 Random graphs
%05C85 Graph algorithms
%68R10 Graph theory (including graph drawing)
%90B15 Network models, stochastic
\end{keyword}

\end{frontmatter}
\linenumbers

\section{Introduction}

The graph dismantling problem pursues for the minimal set of vertices in a graph after removing which the remaining graph will break into many connected sub-graphs of sub-extensive size.
For a graph $\mathcal{G}$ contains $N$ vertices and $E$ links, if we move $\rho N$ vertices and their adjacent links away from $\mathcal{G}$, the remaining graph may not be connected and the relative size of the largest connected component (LCC) will decrease to $g(\rho)$.
The study of this problem contains significant meaning in theoretical studies \cite{Karp2010,Albert2002} and in protecting functional systems \cite{Albert2000, PhysRevLett.85.5468,Cohen2001}.
The function of some complex systems, like power grid or road transportation network, is based on their structure and scale, and the breaking of their connectivity will lead to the function failure.
What is more, when we are facing a spreading epidemic, one way to confine the infection in a small region is vaccinating the people who can dismantle the infection network as far as possible \cite{PhysRevX.4.021024, PhysRevLett.86.3200,Guggiola2015}.
For the same reason, this technology can also be used in controlling computer virus or the spreading of information \cite{Altarelli2013,v011a004}.

%algorithm
It has been proved that the dismantling problem belongs to the class of NP-hard, which means, in the general case, we cannot find the most efficient dismantling sequence by a polynomial algorithm. However, many heuristic strategies have been developed based on various graph properties. For example, we can solve this problem by repeating deleting the vertex with the highest degree \cite{PhysRevLett.85.5468} or betweenness \cite{PhysRevE.65.056109}. A better choice is to delete the highest-degree node from the 2-core of the graph instead of from the original one \cite{Zdeborova2016}. Another development of the algorithm in the same category is to consider the node's collective influence, which is the out-degree of a sub-graph centered around a node with radius $l$\cite{Morone2015}.

Recently, Mugisha and Zhou pointed out that this problem relates to the spanning forest problem, which is also refereed as the feedback vertex set (FVS) problem \cite{Mugisha2016}.
Therefore, a network can be dismantled by the following three-steps algorithm:
Find an approximate solution set of the minimum FVS problem and delete all vertices in the set first.
Then continue breaking the forest by the most efficient way until there is only very small trees left.
At last, optimize the attacking set by bringing some nodes back to the dismantled graph which will not lead to the increase of the LCC.
This algorithm is characterized by its perfect performance in giving a very small threshold value $\rho_c$, where $g(\rho)<0.01$ when $\rho>\rho_c$.
However, another characteristic of this algorithm that the value of $g(\rho)$ decreases slowly before its abrupt drop at $\rho_c$ not only hides the intention of network breaking, but also indicates the intermediate states during the dismantling process is not the optimum.
Along the same line, Braunstein  $et. al.$ also developed a three-states min-sum algorithm to decycle and dismantle a graph \cite{Braunstein2016}.

Except the algorithms discussed above trying to solve this problem from the frontal side, dismantling a graph can be regarded as the reverse process of the node explosive percolation (NEP) \cite{Clusella2016}.
Like the edge explosive percolation, which was introduced by Achioptas $et. al.$ to explain the sudden changing of the relative size of the LCC \cite{Achlioptas2009},
NEP deletes all vertices from a graph first and then recovers them back in a certain order.
Because NEP avoids the emergence of the LCC, the reversed order of NEP will dismantle the LCC as quickly as possible \cite{Schneider2012}.
Therefore, any strategy in NEP can also be used to dismantle network.
In a study of immunization problem \cite{Schneider2012}, Schneider $et. al.$ proposed that a dismantling sequence can be built by inserting the removed vertex toward the front of the sequence repeatedly which gives the least contribution to the size of the LCC.
What is more, the authors of \cite{Clusella2016} took advantage of two complementary node percolation strategies and applied them in the region where they are good at separately.
The numerical results exhibited that blending different strategies together is a practical way to enhance the performance of the NEP algorithm.

In the present paper, we introduce a compound algorithm (CA) which combines the BPD algorithm and the NEP algorithm together.
The BPD algorithm has proved itself as an excellent algorithm in finding the minimal attacking set, but there is still a lot of room for improvement during the entire dismantling process.
Fortunately, the NEP algorithm can make this imperfection up by rearranging the first $T$ elements of the dismantling sequence given by the BPD algorithm.
The numerical computation results on various random graphs demonstrate that the CA improves the performance of the BPD algorithm in the region $\rho\in(0,T/N)$ but does not lead to any degeneration at $\rho > T/N$.
Especially, if we set $T=N\rho_c^{\text {BPD}}$, we will have the $\rho_c^{\text {CA}}\leq \rho_c^{\text {BPD}}$, where $\rho_c^{\text {CA}}$ and $\rho_c^{\text {BPD}}$ are the value of $\rho_c$ given by the BPD algorithm and CA, and at the same time, the general performance of the dismantling sequence is enhanced either.

Most of the existed algorithms discussed above focus their attention on when a graph can be dismantled completely, which is measured by the value of $\rho_c$, but overlook the overall performance during the entire dismantling process.
However, in some cases, the intermediate states during the dismantling is equivalent or more important than merely finding the minimal attacking set.
For the most network dismantling behavior, moving vertex from the graph, which means invalidating functional module or vaccinating healthy people, needs considerable time and cost.
Therefore, the vertices in the minimal attacking set cannot be deleted from the graph synchronously, but can be handled one-by-one.
Facing the outbreak of disease, we hope every vaccinating prevents the epidemic spreading in some degree, rather than only when the whole group of people is vaccinated does it work.
In order to evaluate the general performance of a dismantling algorithm, we introduce another benchmark $R$ in the present paper, which is defined as the area under the curve $g(\rho)$ before $N\rho_c$ vertices are deleted.
This benchmark can also be explained as the robustness of a network \cite{Schneider2012}.
An outstanding dismantling algorithm should be good at giving both a small $\rho_c$ and $R$.

%In the next section we define the FVS problem more precisely and introduce some graph concepts. In Section 3 the spin glass model for the FVS problem on undirected graphs is introduced. This spin glass model is analyzed by the replica- symmetric mean field theory in Section 4 and by belief propagation-guided decimation algorithmin Section 5.We conclude our work in Section 6 and discuss some possible extensions.

This paper is organized as follows: In the next section, we will give the definition of the general performance $R$ first. After that, we review two ancestor algorithms used in the CA first and then introduce how to combine them together.
In Sec.~\ref{sec:Res}, we measure the ability of the CA in improving the general performance numerically under the Erd\"os-R\'enyi (ER) graphs, random-regular (RR) graphs, and scale-free (SF) graphs.
Then, in order to increase the computation efficiency of the CA, we study the relationship between the joint point connecting two ancestor algorithms and the general performance of the CA.
The result of this study helps us implement the fast CA, which is a compromise algorithm in the limited computing resources.
At last, we apply this fast CA in all random graph ensembles and a few real networks here.
In the last section, we conclude the CA and discuss the possible extensions.

\section{The algorithms}

For a graph $\mathcal{G}$ with $N$ vertices, a dismantling algorithm try to give a sequence of vertex $(x_1, x_2, \cdots, x_N)$.
After removing the first $\rho N$ elements of the sequence $(x_i)_{i=1}^N$, the relative size of the LCC in the remaining graph will decrease to $g(\rho)$.
In the present paper, two criterions are used to evaluate the performance of a dismantling algorithm.
A wildly applied benchmark is the threshold value $\rho_c$, which is defined as the value of $\rho$ where $g(\rho>\rho_c)<0.01$ \cite{Morone2015,Mugisha2016, Braunstein2016,Clusella2016}.
Moreover, we also consider the general performance of a dismantling algorithm:
\begin{equation}
\label{eq:R}
R = \int_0^{\rho_c} g(\rho) d \rho.
\end{equation}

In order to highlight the advancement of the CA, we define the relative enhancement of the $R$ as
\begin{equation}
r_R=\frac{R^{\text{BPD}}-R^{\text{CA}}}{R^{\text{BPD}}}\times 100\%,
\end{equation}
where $R^{\text{BPD}}$ and $R^{\text{CA}}$ are the value of $R$ given by the BPD algorithm and CA correspondingly.

In the following subsections, we will review the main steps of two ancestor dismantling algorithms briefly first: the BPD and NEP algorithms.
The reader who are interesting to the details and the theories of them can read original references.
At last, we will explain how to implement our CA mixing two ancestor algorithms together.

\subsection{The BPD algorithm for dismantling problem}

Dismantling algorithms based on the FVS attacking strategy, including the BPD algorithm and the algorithm proposed in \cite{Braunstein2016}, are composed of three main steps: Find the minimum FVS of the graph and add the vertices in the FVS to the sequence $(x_i)$ first. After that, the remaining graph contains no circle anymore and the algorithm continues breaking the remaining forest in the most efficient way.
In the last step, optimize the sequence $(x_i)$ by  kicking some vertices out from $(x_i)$ which will not lead to the increase of the LCC.

In the first step, the BPD algorithm translates the FVS problem with global constrain to a spin-glass model containing local constrain only.
For each vertex $i$, it defines a state $A_i\in\{0, i, j\in\partial i\}$,  which means the vertex $i$ is empty, the root of a tree or the child of neighbour vertex $j$ respectively.
Then the long-range constrain of no loop in the graph can be replaced by a series of local constrains on edges:
\begin{eqnarray}
  & & \hspace*{-0.8cm}
  C_{(i,j)}(A_i,A_j)   \equiv  \delta^0_{A_i}\delta^0_{A_j} \nonumber\\
  & &
  +\delta^0_{A_i}\bigl(1-\delta^0_{A_j}-\delta^i_{A_j}\bigr)
  +\delta^0_{A_j}\bigl(1-\delta^0_{A_i}-\delta^j_{A_i}\bigr) \nonumber  \\
  & &
  +\delta^j_{A_i} \bigl(1-\delta^0_{A_j}-\delta^i_{A_j}\bigr)
  + \delta^i_{A_j} \bigl(1-\delta^0_{A_i}-\delta^j_{A_i}\bigr) \; ,
\end{eqnarray}
where $\delta_x^y$ is the Kronecker symbol such that $\delta_x^y=1$ only when $x=y$, otherwise $\delta_x^y=0$.
Therefore, only when $A_i$ and $A_j$ satisfy all five constrains on edge $(i,j)$, $C_{(i,j)}(A_i,A_j)=1$.
If a microscopic configuration $\underline{A}\equiv (A_1,A_2,\cdots,A_N)$ satisfies all constrains in the graph, we refer it as the legitimate configuration and it is also the solution of the spin-glass model.
For a legitimate configuration $\underline{A}$, if we remove all vertices with $A_i = 0$, the remaining graph contains trees and a few cycle-trees only, which is a kind of subgraph with only one loop.
Therefore, the vertices with $A_i = 0$ in the $\underline{A}$ can be regarded as forming a FVS approximately.
The energy of the $\underline{A}$, corresponding to the size of the FVS, is defined as the number of vertices with $A_i = 0$.

In order to find the ground state of this spin-glass model, the BPD algorithm follows the standard steps of the cavity method to build belief-propagation (BP) equations by defining a pair of messages (probability distribution) on each edge $(i,j)$: $q^A_{i\rightarrow j}$ and $q^A_{j\rightarrow i}$, where $q^A_{i\rightarrow j}$ is the probability of vertex $i$ taking the state $A$ in the absence of vertex $j$.
Considered the constrains on edges, all messages in the graph $\mathcal{G}$ should fulfill the following the BP equations:
\begin{subequations}
  \label{eq:FVS_BP}
  \begin{align}
    q^0_{i\rightarrow j}& =
    \frac{e^{-\beta}}{z_{i\rightarrow j}} \; , \label{eq:BPe} \\
    q^i_{i\rightarrow j} & =
    \frac{1}{z_{i\rightarrow j}}
    \prod_{k \in \partial i\backslash j} \bigl[
      q^0_{k\rightarrow i}+q^k_{k \rightarrow i}\bigr]
    \; , \label{eq:BPr} \\
    q^l_{i\rightarrow j} & =  \frac{(1-q^0_{l\rightarrow i})}{z_{i\rightarrow j}}
    \prod_{k \in \partial i\backslash j,l} \bigl[
      q^0_{k\rightarrow i}+q^k_{k \rightarrow i}\bigr]
    \; , \quad  (l\in \partial i \backslash j)
    \label{eq:BPo}
  \end{align}
\end{subequations}
where $\beta$ is the inverse temperature and the normalization factor $z_{i\rightarrow j}$ is
\begin{equation}
\label{eq:FVS_zij}
  z_{i\rightarrow j}\equiv e^{-\beta}+ \Bigl[1+\sum_{k \in \partial i\backslash j}
    \frac{1-q^0_{k\rightarrow i}}{q^0_{k\rightarrow i}+q^k_{k \rightarrow i}} \Bigr]
  \prod_{l \in \partial i\backslash j}\bigl[ q^0_{l\rightarrow i}+q^l_{l \rightarrow i}
    \bigr] \; .
\end{equation}
Therefore, the probability of vertex $i$ taking the state 0 is,
\begin{equation}
\label{eq:FVS_q0i}
q_i^0=\bigg[e^{-\beta}+ \Bigl[ 1+\sum_{k \in \partial i} \frac{1-q^0_{k\rightarrow i}}{q^0_{k\rightarrow i}+q^k_{k \rightarrow i}}    \Bigr]  \prod_{j \in \partial i}\bigl[q^0_{j\rightarrow i}+q^j_{j \rightarrow i}\bigr]\bigg]^{-1}.
\end{equation}

The BPD algorithm can figure out which vertex should take the state $0$ by repeating the following two steps until there are only trees or cycle-trees left in the remaining graph:
Iterate the Eq.~\ref{eq:FVS_BP} with a very large $\beta$ enough times and then compute $q_i^0$ for each vertex by Eq.~\ref{eq:FVS_q0i}.
Delete the vertex with the largest $q_i^0$ from the graph, and add this vertex to the tail of the sequence $(x_i)$.

Now, we can continue breaking cycle-trees by attacking anyone vertex on the cycle and push it into the back of the attacking sequence $(x_i)$.
In the following discussion, we use $\mathcal{T}_\alpha$ to denote tree $\alpha$ in the remaining forest $\mathcal{F}$.

The second stage of the BPD algorithm will break the $\mathcal{F}$ as quickly as possible. To achieve this purpose, it defines another pair of messages for each edge $(i,j)\in\mathcal{F}$: $n_{i\rightarrow j}$ and $n_{j\rightarrow i}$. $n_{i\rightarrow j}$ saves the size of the tree containing vertex $i$ in the case of cutting edge $(i,j)$.
Then we can write the iteration equations of these messages:
\begin{equation}
\label{eq:nij}
n_{i\rightarrow j}=1+\sum_{k\in \partial i\setminus j} n_{k \rightarrow i}.
\end{equation}
These equations can be solved easily by starting them from all leaf vertices and then propagating the messages to the entire tree gradually.

If the vertex $i\in\mathcal{T}_\alpha$ is attacked, $\mathcal{T}_\alpha$ will break to a few smaller trees with the size $\{n_{j\rightarrow i}\}_{j\in \partial i}$, and the size of the largest one in them would be
\begin{equation}
n_i=\max\limits_{j \in \partial i} n_{j \rightarrow i}.
\end{equation}
Therefore, the most efficient way of breaking $\mathcal{T}_\alpha$ is to attack the vertex with the smallest $n_i$.
What is more, in order to decrease the size of the LCC in $\mathcal{F}$ quickly, we should dismantle the largest tree in $\mathcal{F}$ firstly.
Using these strategies, a forest with giant trees will break to one with numerical tiny trees in the size no larger than the threshold value $0.01N$.

Now, the size of the LCC in the remaining graph will be equal to or just smaller than $0.01N$, but there will be a mass of subgraphs whose size is much smaller than the threshold value. In order to optimize the attacking sequence, in the last step of the BPD algorithm, the vertices which could bring small loops but which will not lead to the enlargement of the size of the LCC can be kicked out from the attacking sequence $(x_i)$.

The BPD algorithm is rich threatening. Because the attacker hides the attacking attention during the dismantling process. The structure of a graph will be trimmed to a very fragile one gradually in the first stage and then the attacker dismantles the remaining graph quickly in the second. In that case, people will not realize they are facing dismantling attack at the beginning and it will be too late to prevent the cascading failure when the attacking behavior is already in the second stage. What is more, the BPD algorithm also behaves well in giving a small $\rho_c$, which means it is very efficient in breaking the original graph to small pieces.

\subsection{The NEP algorithm for dismantling problem}
\label{sec:NEP}

Different from most traditional dismantling algorithms, the NEP algorithm solves this problem using a backward thinking.
It starts from a dismantled graph and then recovers the vertex with attached edges back.
%This problem can be answered by defining a score for each removed vertex.
%In each step, we compute the score of all deleted vertices first.
%Then the vertex with the smallest score should be recovered to the graph in priority.
%At the same time, we also push this vertex to the front of the sequence $(x_i)$.
During the recovering process, separated small subgraphs merge to a bigger one which could accompany with the increase of the size of the LCC.
The starting point of the NEP is just to avoid the increase of the LCC as far as possible.
Therefore, the NEP dismantling algorithm gives an attacking sequence reducing the size of the LCC as quickly as possible.
In order to achieve this target, a score $\sigma_i$ measuring the ability of vertex $i$ merging surrounding subgraphs is defined for each removed vertex
and the vertex with the minimal $\sigma_i$ should be recovered to the dismantled graph with priority.
At the same time, we push the vertex $i$ to the head of the sequence $(x_i)$.
After recovering the last removed vertex, we assemble the original graph back and generate a complete dismantling sequence $(x_i)$.

From the description above, we can see the performance of the NEP algorithm depends on the definition of the vertex score, which should consider structure features of the surrounding subgraphs.
For example, in the simplest way, $\sigma_i$ can be defined as the size of the new subgraph generated by adding vertex $i$ to the remaining graph \cite{Schneider2012}, which is denoted as the definition 1 (D1) in this paper.
This definition prevents the increase of $g$ in the present, which might lead to an unexpected increase of $g$ in the future.
In order to overcome the drawback of D1, we also study the second score definition (D2) from \cite{Clusella2016}:
\begin{equation}
\sigma_i=|\mathcal{N}_i|+\epsilon |\mathcal{C}_{i,2}|,
\end{equation}
where $|\mathcal{N}_i|$ is the number of neighbour subgraphs connected with vertex $i$, $|\mathcal{C}_{i,2}|$ is the size of the second largest neighbour subgraphs of vertex $i$, $\epsilon$ is a very small positive number here. Therefore, the NEP algorithm with D2 prefers to recover the vertex with few neighbour subgraphs. If more than one vertex has the minimal $|\mathcal{N}_i|$, it selects the vertex with the smallest $|\mathcal{C}_{i,2}|$.

\subsection{The compound algorithm mixing BPD and NEP algorithms}

%In the present paper, I take the BPD algorithm as an example to explain how to improve the performance an already existed dismantling algorithm by NEP.

There is a useful property in dismantling problem that after a few vertices are removed from a graph, the structure of the remaining graph is merely decided by the set of the removed vertices and does not depended on the order they moved.
As described above, we can see the BPD algorithm solves a dismantling problem from the frontal side and the NEP algorithm handles it following the reversed way.
Therefore, the NEP algorithm can rearrange the first $T$ elements of a dismantling sequence, which might improve the performance of the original sequence in the first $T$ steps without any negative influence in the rest part.
Specifically, we can take advantage of the NEP algorithm to optimize a dismantling sequence generated by the BPD algorithm. This strategy generates a CA with the merit of both algorithms.

The CA introduced here is implemented in this way:
First, find a dismantling sequence $(x_i)_{i=1}^N$ by BPD algorithm.
Then, delete the first $T$ vertices in $(x_i)_{i=1}^N$ from the original graph and pop these vertices out from $(x_i)_{i=1}^N$ simultaneously.
Now, we have the tail part of the original sequence $(x_i)_{i=T+1}^N$ and a partially dismantled graph $\mathcal{G}_T$.
%????
At last, use the NEP algorithm with a specific score definition to rearrange the order of the deleted $T$ vertices and rebuild the head part of $(x_i)_{i=1}^N$.

The last question in the CA is how to select the joint point $T$ where two algorithms are connected.
Considering the benchmarks $\rho_c$ and $R$ are used to evaluate the quality of a dismantling algorithm in the present paper, we should choose the joint point where one or both benchmarks can be optimized.
However, in most cases, $\rho_c$ and $R$ cannot be minimized simultaneously and we can only focus on one of them or give consideration to a synthesized benchmark.
In the present paper, we would like to design an algorithm which can dismantle a graph efficiently during the entire process, which is evaluated by the $R$.
Therefore, two ancestor algorithms should be connected at the point where the value of $R$ can be minimized.
We should also declare that the behavior of merely optimizing $R$ leads to the degeneration of the another benchmark $\rho_c$.
When facing a problem in which dismantling a graph to small pieces quickly is the priority, we can also choose $T$ with the minimal $R$ on the condition of $\rho^{\text{CA}}_c\leq\rho^{\text{BPD}}_c$.

\section{Results}
\label{sec:Res}
In this section, we evaluate the performance of the CA on three different random graph ensembles: ER graph, RR graph and SF graph.
Without being specific, the results in the following discussion are obtained by averaging over 16 different graph instances with $N=2^{16}$.
%Besides that, we also apply the fast CA on these random graph ensembles and some real network instances.
By comparing the benchmarks of two ancestor algorithms and that of the CA, we can say that the CA inherits the advantage from both of its ancestors.

\begin{figure}
\begin{center}
\includegraphics[width=0.8\textwidth]{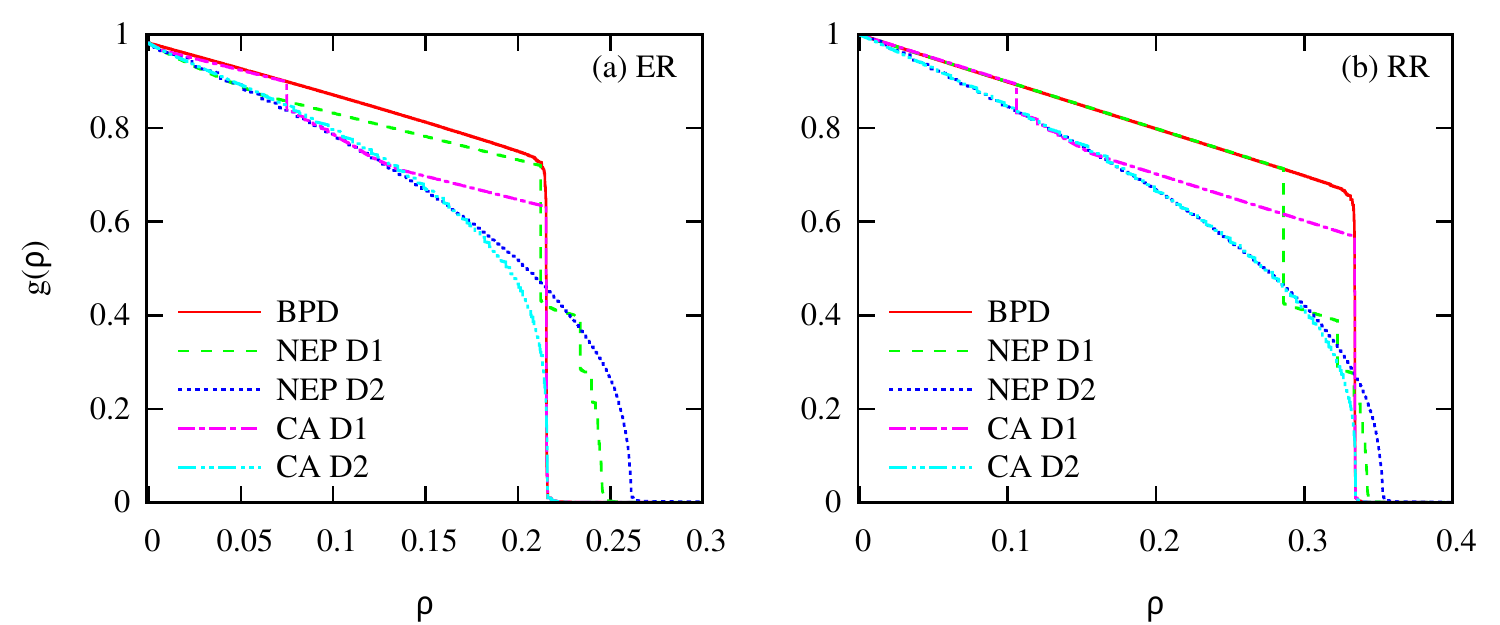}
\caption{
\label{fig:fig1}
The relative size of the LCC $g$ in a ER graph with $c=4.0$ (a) and a RR graph with $K=4$ (b) with the fraction of removed vertex $\rho$.
In these figures, we present the result of BPD algorithm (solid line), the NEP algorithm with two different vertex score definitions(dash line and dot line), and the CA with these vertex score definitions (dash-dot line and dash-dot-dot line).
}
\end{center}
\end{figure}

Figure 1 presents the relationship between the relative size of the LCC $g$ and that of the removed vertices $\rho$, given by various algorithms on one ER graph instance with average degree c=4.0 and one RR graph instance with degree K=4.
We can observe that the BPD algorithm gives a very small $\rho_c^{\text{BPD}}$, which is 0.2162 in the ER graph and 0.3346 in the RR graph.
As comparison, the NEP algorithms with D1 and D2 give $\rho_c^{\text{NEP D1}}=0.2468$ and $\rho_c^{\text{NEP D2}}=0.2625$ in ER graph and 0.3435, 0.3538 in RR graph separatively.
Although the NEP algorithm is not good at giving a small dismantling set, it is superior to the BPD algorithm in the region $\rho<\rho_c^{\text{BPD}}$.
What is more, sometimes, the NEP algorithm is better than the BPD algorithm in giving a smaller $R$.
For example, in the case of this ER and RR graph, the $R$ of the BPD algorithm are 0.1852 and 0.2777 respectively, and those of the NEP algorithm with D2 are 0.1773 and 0.2397.
Based on these facts, a simple thought is applying two complementary algorithms separately in the region where they are suitable for, and that is just the starting point of our CA.
When applying our CA with one of two score definitions in this ER graph, we find the general performance $R$ is improved remarkably compared with that of the BPD algorithm and even with that of the NEP algorithm, which reaches 0.1708 and 0.1611 for the CA with D1 and D2.
In the case of this RR graph, the CA also reduces the $R$ of the BPD algorithm: $R$ of the CA with D1 and D2 reaches 0.2570 and 0.2351 correspondingly.
We should also declare that, the decrease of the $R$ in CA might be accompanied with the degeneration of the $\rho_c$.
However, the price of improving the general performance in CA is negligible.
Because we only observe a very small degeneration in this RR instance for the CA with D2 and the relative increase of the $\rho_c^{\text{CA}}$ is $0.03\%$ compared with $\rho_c^{\text{BPD}}$.
%0.000273648

%For the CA, there is no much room for improvement in $\rho_c$, but that in $R$ is obvious:
%In the case of ER graph, and when OT is $R$, $r_R=9.0\%$ and $15.4\%$ for D1 and D2 respectively compared with that of the BPD algorithm.
%Even when $\rho_c$ is the OT, there is still a decreasing in $R$ with $r_R=8.0\%$ and $15.0\%$ for D1 and D2 respectively.
%We can also observe a similar result in the case of the RR graph.
%On the contrary, for D1 and D2 respectively, the improvement in $\rho_c$ is merely $r_{\rho_c}=0.61\%$ and $r_{\rho_c}=0.15\%$ in ER graph and $r_{\rho_c}=1.52\%$ and $r_{\rho_c}=0.27\%$ in RR graph.

\begin{figure}
\begin{center}
\includegraphics[width=1.0\textwidth]{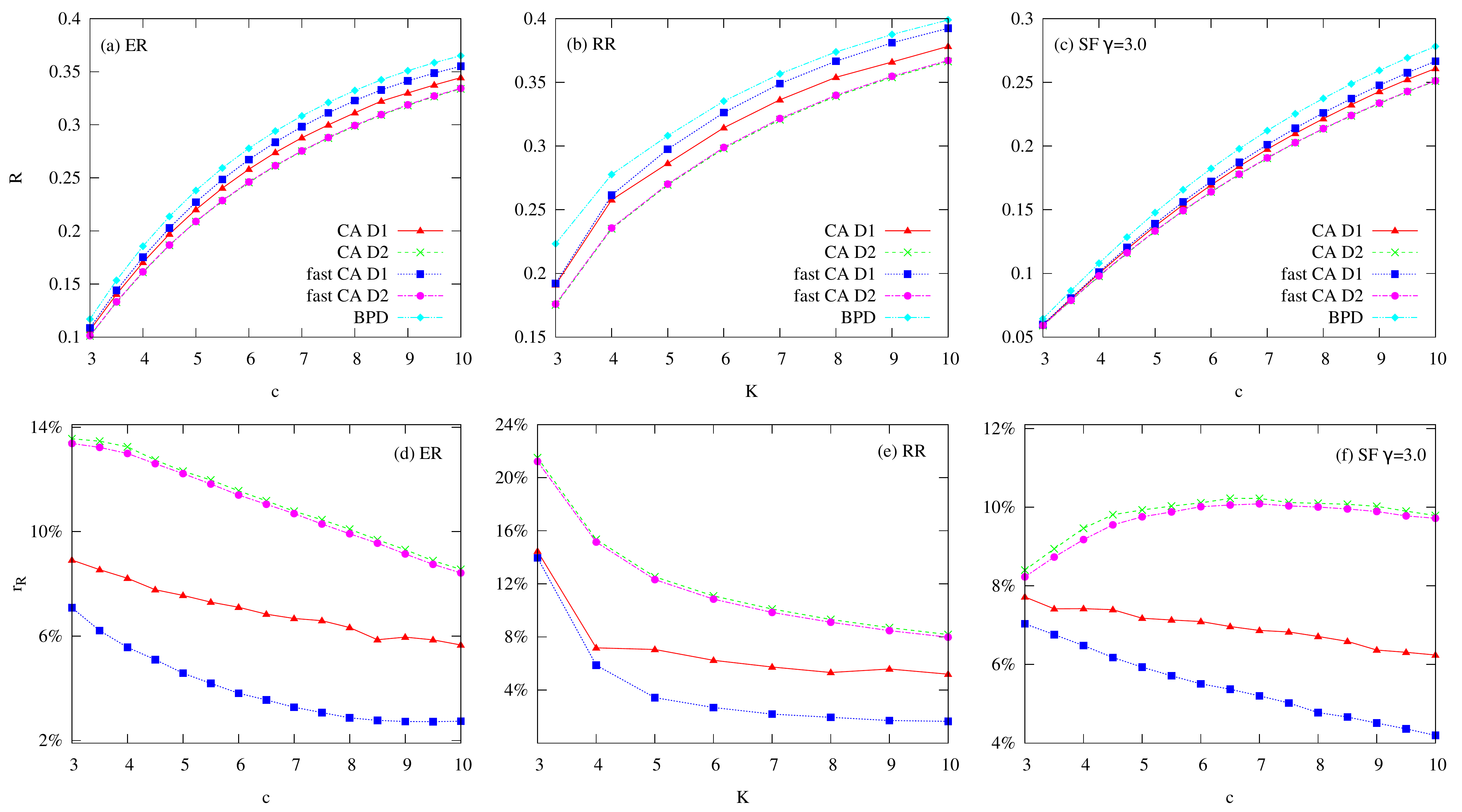}
\caption{
\label{fig:fig2}
The general performance $R$ for ER graph in the function of mean degree $c$ (a), RR graph in the function of degree $K$ (b) and SF graph with degree decay exponent $\gamma=3.0$ in the function of mean degree $c$ (c).
In these figures, we present the results of BPD algorithm, CA with D1 and D2 and the fast CA with D1 and D2.
Figure (d), (e) and (f) are the relative enhancement $r_R$ of the CA with D1 and D2 and the fast CA with D1 and D2 in three corresponding random graph ensembles.
}
\end{center}
\end{figure}

In order to highlight the advantage of the CA, we investigate the value of $R$ and $r_R$ of all these algorithms in ER, RR and SF graphs ($\gamma=3.0$) with various mean degree systematically, and then present all results in Fig.~\ref{fig:fig2}.
In the present paper, we generate the SF networks by a static method explained in \cite{PhysRevLett.87.278701}.
For all algorithms, with the increase of the connectivity, it becomes harder and harder to dismantle a graph, and the effect of the CA also becomes weaker.
However, there still exist obvious enhancement in the CA.
Especially for the CA with D2, the relative improvement $r_R$ is larger than $8\%$ in all tested instances.
We also find that, for all these random graph ensembles, the CA with D2 is more suitable than that with D1 in improving the general performance.

\begin{figure}
\begin{center}
\includegraphics[width=0.8\textwidth]{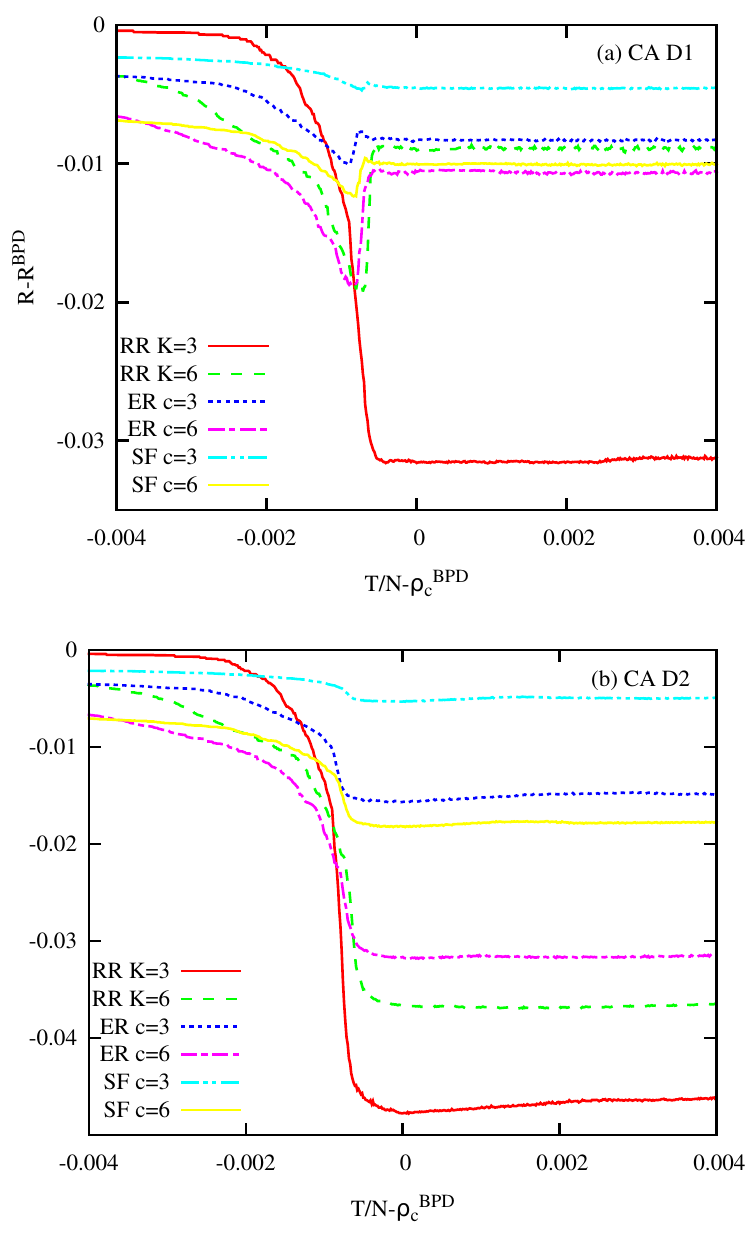}
\caption{
\label{fig:fig3}
The comparison of the general performance $R$ between the BPD algorithm and the CA with D1 (a) and D2 (b) in the function of the joint point $T$ on ER, RR and SF graphs with different degrees. The degree decay exponent $\gamma=3.0$ for the SF graph.
}
\end{center}
\end{figure}

\begin{figure}
\begin{center}
\includegraphics[width=0.4\textwidth]{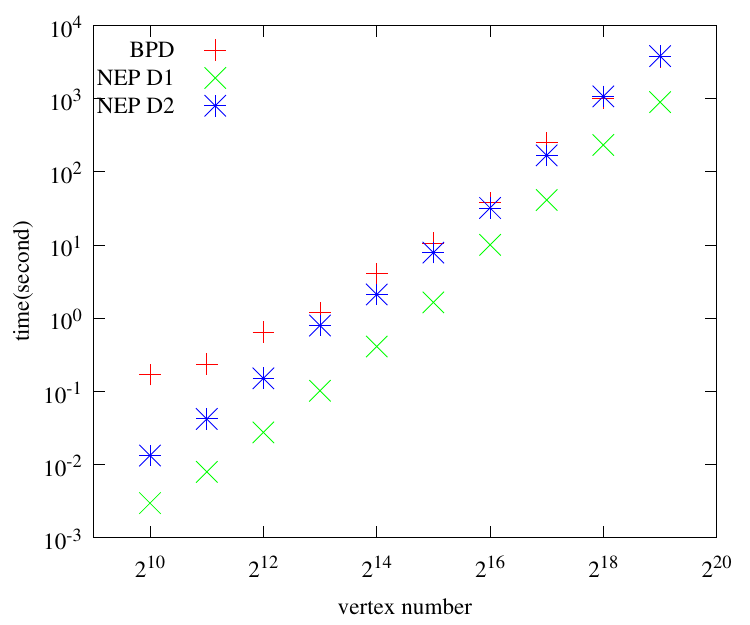}
\caption{
\label{fig:fig4}
The computation time of two ancestor algorithms (the BPD and NEP algorithm) in the fast CA on ER graph with $c=3.0$ and vertices number from $N=2^{10}$ to $2^{19}$ (Intel Xeon E5450, 3.00GHz, 2GB memory).
}
\end{center}
\end{figure}

Based on the description of the CA above, it seems like that we should repeat the NEP algorithm $N$ times to find the best joint point where $R$ can be optimized.
In that case, the CA will be at least $N$ times slower than the NEP algorithm.
Considered the cost of the computation time, the price of the merit in CA seems to be too expensive.
In order to prevent the expanding of the computation cost, it is necessary to reduce the searching space of joint point $T$.
For this purpose, we also study the behavior of $R$ with the position of joint point $T$ and present the results in Fig.~\ref{fig:fig3}.
%Actually, when $T<N\rho_c^{\text{BPD}}$, $\rho_c$ is decided by the BPD algorithm, so the CA cannot improve the $\rho_c$ in this region.
%In the region $T\leq N\rho_c^{\text{BPD}}$, the average of $\rho_c$ suddenly drops to its minimal first and then increases with $T$ gradually.
At first, $R$ of the CA with D1 and D2 decrease gradually together with increasing $T$.
Except for the RR graph with K=3, before the graph broken completely, where $T<N\rho_c^{\text{BPD}}$, $R$ of the CA with D1 starts to rebound and raises to a stable stage.
On the other hand, $R$ of the CA with D2 keeps falling until reaches its minimal around $N\rho_c^{\text{BPD}}$.
Therefore, it is not necessary for us to compute and compare the value of $R$ through the entire searching space $T\in[0,N]$.
We should only focus our attention in a very narrow region around $N\rho_c^{\text{BPD}}$.
For example, the most efficiency preference choice is using $T=N\rho_c^{\text{BPD}}$ directly and we denote it as the fast CA in the present paper.

Then we also study the general performance $R$ and $r_R$ of the fast CA and present the results in Fig.~\ref{fig:fig2} either.
It is easy to understand that the results of the fast CA is worse than that of the ordinary CA searching the whole space $T\in [0,N]$.
However, the fast CA brings another benefit: The $\rho_c$ will not degenerate anymore.
% although there is no obvious difference between two results for the CA with D2.
In the case of CA with D1, the difference of $r_R$ between the fast CA and the ordinary CA is obvious, so it is worth searching the best $T$ in a region before $N\rho_c^{\text{BPD}}$.
On the other hand, for the CA with D2, considering that $R$ reaches its minimal before $N\rho_c^{\text{BPD}}$ and expanding the searching space does not improve the performance of the algorithm conspicuously, we can just select the $T=N\rho_c^{\text{BPD}}$ and use fast CA directly.

In order to figure out the computation complexity of the fast CA, we investigate the computation time of the fast CA with the size of the graph.
As the fast CA is composed of BPD and NEP algorithms, we present the computation time of two parts separately in Fig.~\ref{fig:fig4}.
Generally speaking, for graphs with the same average degree $c$ and various vertex numbers $N$, the computation complexity of the NEP stage is in the same order as that of the BPD stage, which is $\mathcal{O} (N \ln N)$.
Therefore, the computation complexity of the fast CA is also in the same order as that of the BPD algorithm.

%As the discussion above, the characters of the CA are decided by the score definitions used in the algorithm, and CA with D1 and D2 are good at optimizing $\rho_c$ and $R$ separately.
%The authors in Ref.~\cite{Clusella2016} have mixed two score definitions together to improve the quality of the NEP algorithm.
%In this paper, we follow the example of Ref.~\cite{Clusella2016} and confirm that introducing two score definitions to our CA can make it perform well in both aspects.
%The general line of thought of our mixing strategy is that: Use the CA with D1 to fix the value of $\rho_c$ first and then apply CA with D2 to trim the intermediate states during the dismantling at $T=N\rho_c$.
%We use CA mixed 2 score definitions to compute the value of $r_{\rho}$ and $r_R$ for both ER and RR graphs and present the results in Fig.~\ref{fig:fig4}. Each point in the graph shows the value of $r_{\rho}$ and $r_R$ of an algorithm in a graph ensemble, and we use the black dash lines connect the results of the same graph ensemble.
%From this figure, we can also see that the results of D1 get together at the right bottom and the blue points distribute in the left top.
%The black dash lines are vertical and nearly horizontal, which means the CAM2D improve the $\rho_c$ and $R$ at the same time.

At last, we apply the fast CA to dismantle a set of real-world networks to small components with the size smaller than 0.01N.
Different from the random graph ensembles, all these real-world networks contain plenty of communities, local loops, and hierarchical levels.
We list the value of $R$ of the BPD algorithm and that of the fast CA with D1 and D2 in table~\ref{tab:tab1}.
The fast CA improves the general performance of the BPD algorithm remarkably from $r_R=8.0\%$(fast CA with D1 in Citation network) to $r_R=81.9\%$(fast CA with D1 in the RoadTX network).
Difference from the case of the random graph ensemble, for the real-world networks, the fast CA with D1 seems to be more suitable for trimming the dismantling sequence than that with D2:
Although the fast CA with D1 gives a better dismantling sequence than that with D2 only in 7 of 12 real-world networks, which is not overwhelming, the average relative improvement of the fast CA with D1 $\overline{r}_R=30.1\%$ is almost twice of that with D2 $\overline{r}_R=17.2\%$.

\begin{table}[!hpb]
\centering
\caption{\label{tab:tab1}Comparative results of the BPD algorithm and the fast CA with two vertex score definitions on a set of real-world network instances.
N and E are the number of vertices and links of each network respectively.
%The number of vertex $N\rho_c$ removed by the BPD algorithm is listed in the 4th column, and
The general performance $R$ of the BPD algorithm and fast CA with D1 and D2 are listed in the 4th, 5th, and 6th column correspondingly. We use boldface to highlight the minimum $R$ in all three algorithms.
}
\begin{tabular}{lrrrrrccc}
\cline{1-9}
Network & N & E & &\multicolumn{1}{c}{BPD}& &\multicolumn{1}{c}{fast CA D1}& &\multicolumn{1}{c}{fast CA D2}\\
\cline{1-9}
RoadEU\cite{Subelj2011}    &1177     &1417      &     &0.04550&&$\bf{0.01950}$&&0.04176\\
PPI \cite{Bu2003}       &2361     &6646       &    &0.09230&&$\bf{0.07889}$&&0.08089\\
Grid\cite{Watts1998}      &4941     &6594        &   &0.03547&&$\bf{0.009722}$&&0.02978\\
IntNet1\cite{Leskovec2005Graphs}   &6474     &12 572      &   &0.01226&&$\bf{0.009195}$&&0.01040\\
Authors\cite{Leskovec:2007:GED:1217299.1217301}   &23 133   &93 439      & &0.08767&&0.07630&&$\bf{0.07546}$\\
%citHepPh
Citation \cite{Leskovec2005Graphs} &34 546   &420 877     &&0.2930&&0.2695&&$\bf{0.2601}$\\
%p2pGnutella
P2P\cite{Leskovec2005Graphs}       &62 586   &147 892      &&0.1155&&0.1047&&$\bf{0.1038}$\\
%locGowalla
Friend\cite{Leskovec2005Graphs}    &196 591  &950 327     &&0.1028&&0.09182&&$\bf{0.08926}$\\
%emailEuAll
Email \cite{Leskovec:2007:GED:1217299.1217301}    &265 214  &364 481      & &0.001293&&0.0008334&&$\bf{0.0008297}$\\
WebPage\cite{Leskovec2009}   &875 713  &4 322 051   &&0.04937&&$\bf{0.04103}$&&0.04334\\
RoadTX\cite{Leskovec2009}    &1 379 917&1 921 660   &&0.01271&&$\bf{0.002304}$&&0.007011\\
intNet2 \cite{Leskovec2005Graphs}   &1 696 415&11 095 298  &&0.03715&&$\bf{0.03082}$&&0.03211\\
\cline{1-9}
\end{tabular}
\end{table}

\section{Conclusion}

In order to design an algorithm which can dismantle a graph efficiently during the entire dismantling process, we combine two excellent algorithms together: the BPD and NEP algorithm.
The BPD algorithm is good at giving a very small dismantling set, which will break the original graph to subgraphs with the size smaller than the $1\%$ of the original graph, and the NEP algorithm is more efficient in the intermediate states during the dismantling process.
Therefore, we combine them together to implement a new CA.
Large-scale numerical computations on ER, RR, and SF networks and some real-world networks reveal that
although the CA gives a dismantling threshold $\rho_c$ as the same as or a little larger than that of the BPD algorithm, the CA improves the overall performance of the BPD algorithm during the entire dismantling process.
However, there exist a serious drawback in the CA: In order to search the best joint point $T \in [0,N]$, it is in the higher level of the computation complexity than that of the NEP algorithm.
Fortunately, we find that setting $T=N\rho_c^{\text{BPD}}$ directly is a good compromise between finding a better dismantling sequence and saving computation resources.
We call the CA with $T=N\rho_c^{\text{BPD}}$ as the fast CA, and it keeps the same computation complexity with the BPD.
What is more, the improvement of the algorithm efficiency is not in the cost of an obvious degeneracy of final result anymore.

The CA discussed in the present paper mixes two specific algorithms: the BPD and NEP algorithm.
Actually, in the same thinking, we can also design other CA by mixing any number dismantling algorithms.
As long as we can find proper joint points, the CA will inherit multiple advantages from all it ancestors and will not be in a higher complexity class than the slowest ancestor algorithm.
In short, mixing different algorithms together is a practicable strategy to improve the performance of an existed one.

\section*{Acknowledgement}

The author thanks Pan Zhang and Salomon Mugisha for helpful discussion.
The computations in this paper are carried out in the HPC Cluster of ITP-CAS.
This work is supported by the 2016 Basic Operational Outlays for the Research Activities of Centric University, Civil Aviation University of China (Grant no. 3122016L010).

\bibliography{bibfile}{}
\bibliographystyle{unsrt}
\end{document}